\documentclass[11pt]{article}
\usepackage{iap2000,epsfig}

\def\Journal#1#2#3#4{{#1} {\bf #2}, #3 (#4)}

\def\aj{\em Astronomical Journal}
\def\aap{\em Astronomy \& Astrophysics}

\def\apj{\em Astrophysical Journal}

\def\mnras{\em Monthly Notices of the Royal Astron.\ Soc.}
\def\nat{\em Nature}

\def\mco{\multicolumn}
\def\be{\begin{equation}}
\def\ee{\end{equation}}
\def\bea{\begin{eqnarray}}
\def\eea{\end{eqnarray}}
\def\ra{R_{\rm A}}
\def\kms{\mbox{km\,s}^{-1}}

\def\df{\mbox{DEF}}
\def\fdef{F_{\rm DEF}}
\def\degr{^\circ}
\def\arcmin{\prime}
\def\arcsec{\prime\prime}
\begin{document}
\vspace*{4cm}

\title{LOOKING FOR CLUES TO THE NATURE OF HI DEFICIENCY IN CLUSTER
SPIRALS~\footnote{For further details, see J.M.\ Solanes, A.\ Manrique,
C.\ Garc\'\i a-G\'omez, G.\ Gonz\'alez-Casado, R.\ Giovanelli, and
M.P.\ Haynes, {\em Astrophysical Journal}, in press (2000);
astro-ph/0007402.}}

\author{J.M. SOLANES}

\address{Departament d'Enginyeria Inform\`atica i Matem\`atiques,
Universitat Rovira i Virgili.\\Carretera de Salou, s/n;
E--43006~Tarragona, Spain}

\maketitle \abstracts{We derive the atomic hydrogen content for 1900
spirals in the fields of eighteen nearby clusters. By comparing the HI
deficiency distributions of the galaxies inside and outside one Abell
radius ($\ra$) of the center of each region, we find that two thirds of
the clusters in our sample show a dearth of neutral gas in their
interiors. Possible connections between the gaseous deficiency and the
characteristics of both the underlying galaxies and their environment
are investigated in order to gain insight into the cause of HI
depletion. In the clusters in which neutral gas deficiency is
pronounced, we see clear indications that the amount of depletion is
related to the morphology of the galaxies: early-type and, probably,
dwarf spirals are more easily emptied of gas than the intermediate
Sbc--Sc types. Gas contents below one tenth, and even one hundredth, of
the expected value have been measured, implying that gas removal must
be very efficient. Our 21-cm data also show that in the HI-deficient
clusters the proportion of gas-poor spirals decreases continuously
towards the outskirts of these systems, the zone of significant
deficiency reaching as far out as $2\ra$. In an independent analysis of
the Virgo cluster, we find suggestive indications that gas losses are
driven by the interaction of the disks with the inner hot intracluster
gas around M87. We also report evidence that gas-poor spirals follow
more radial orbits than those of the gas-rich objects. We conclude that
ISM-IGM hydrodynamic effects appear as the most plausible cause of HI
removal.}

\section{The Cluster Sample}\label{sample}

The galaxies used in the present study have been extracted from the
all-sky database of nearby galaxies maintained by R.\ Giovanelli and
M.P.\ Haynes known as the Arecibo General Catalog (AGC). To be assigned
to a given cluster field, a galaxy must lie within a projected distance
of $5\ra$, i.e.\ within 7.5\,$h^{-1}$ Mpc, of the cluster center and
have a radial velocity which is separated from the recessional velocity
of the cluster no more than $\sim 3$ times its average velocity
dispersion. Since we are especially interested in the central portions
of clusters where environmental influences are strongest, we have only
selected those clusters with at least 10 galaxies (of types
Sa--Sdm/Irr) with good HI detections located within the innermost
$1\ra$ (1.5\,$h^{-1}$ Mpc) radius circle.

HI deficiencies for individual galaxies have been quantified by means
of a parameter $\df$ that compares, the observed HI mass, $h^2M_{\rm
HI}^{\rm obs}$, in solar units, with the value expected from an
isolated (i.e.\ free from external influences) galaxy of the same
morphological type, $T^{\rm obs}$, and optical linear diameter,
$hD_{\rm opt}^{\rm obs}$, expressed in kpc (for details, see Haynes and
Giovanelli~\cite{HG84}): \be \df=\;\langle\log M_{\rm HI}(T^{\rm
obs},D_{\rm opt}^{\rm obs})\rangle-\log M_{\rm HI}^{\rm obs}\;,
\label{def2} 
\ee 
so positive values of $\df$ indicate HI deficiency. For the
expectation value of the (logarithm of the) HI mass, we use the maximum
likelihood linear regressions of $\log(h^2M_{\rm HI})$ on $\log(hD_{\rm
opt})$ inferred from isolated galaxies by Solanes {\it et
al.}~\cite{SGH96} Because the standards of normalcy for the HI content
are well defined only for the giant spiral population (Sa--Sc), we have
excluded from the present study earlier Hubble types, as well as all
galaxies unclassified or with peculiar or very disturbed
morphologies. Nonetheless, HI mass contents for the few HI-rich Sd's,
Sdm's, and Magellanic-type irregulars included in our samples have been
calculated from the relationship inferred for the Sc's.

\begin{table}[t]
\caption{Cluster Fields \label{tab:tab1}}
\vspace{0.1cm}
\begin{center}
\small
\begin{tabular}{|lrclcccrcccr|}
\hline
\mco{1}{|c}{} & \mco{3}{c}{R.A. (1950)} &
\mco{3}{c}{Dec. (1950)} & \mco{1}{c}{Velocity Filter} & 
\mco{1}{c}{$\ra$} & \mco{2}{c}{\# of Galaxies} & 
\mco{1}{c|}{}\\ 
\mco{1}{|c}{Name} & \mco{1}{c}{(h)} &
\mco{1}{c}{(m)} & \mco{1}{c}{(s)} & 
\mco{1}{c}{($\degr$)} & \mco{1}{c}{($\arcmin$)} & 
\mco{1}{c}{($\arcsec$)} & \mco{1}{c}{($\kms$)} & 
\mco{1}{c}{($\deg$)} & \mco{1}{c}{$r\le 1\ra$} & 
\mco{1}{l}{$r\le 5\ra\!$\footnotemark} & 
\mco{1}{c|}{$P_{\rm KS}$}\\
\hline
Pisces &00&59&54.0& $+$30 & 02 & 00 & 3500$\div$7500$\ $ & 1.88 & $\ $14 & 155 & 0.316\\ 
N507 &01&24&26.8& $+$34 & 03 & 35 & 3500$\div$7500$\ $ & 1.83 & $\ $19 & 124 & 0.497\\ 
A262 &01&49&49.9& $+$35 & 53 & 50 & 3000$\div$7000$\ $ & 1.90 & $\ $35 & 168 & 0.002\\ 
A397 &02&53&51.2& $+$15 & 41 & 35 & 8500$\div$11000$\ $ & 0.95 & $\ $10 & $\ $26 & 0.003\\
A400 &02&55&00.0& $+$05 & 48 & 25 & 5000$\div$9000$\ $ & 1.28 & $\ $33 & 100 & 0.057\\ 
A426 &03&16&30.0& $+$41 & 20 & 00 & 2000$\div$9000$\ $ & 1.70 & $\ $12 & $\ $99 & 0.008\\ 
A539 &05&13&55.2& $+$06 & 23 & 16 & 6500$\div$10500$\ $ & 1.03 & $\ $22 & $\ $38 & 0.909\\
Cancer &08&17&00.0& $+$21 & 11 & 00 & 2500$\div$7000$\ $ & 1.75 & $\ $17 & $\ $83 & 0.581\\
A779 &09&16&44.3& $+$33 & 57 & 18 & 4500$\div$9000$\ $ & 1.25 & $\ $11 & $\ $28 & 0.012\\ 
A1060 &10&34&27.7& $-$27 & 16 & 26 & 2000$\div$5500$\ $ & 2.16 & $\ $20 & $\ $96 & 0.060\\
A1367 &11&42&04.6& $+$19 & 59 & 14 & 4000$\div$9000$\ $ & 1.32 & $\ $28 & 100 & $<$0.001\\
Virgo &12&28&18.0& $+$12 & 40 & 00 & $-$500$\div$2700$\ $ & 7.23 & 218 & 426 & $<$0.001\\
A3526 &12&46&06.0& $-$41 & 02 & 00 & 1400$\div$4500$\ $ & 2.54 & $\ $12 & $\ $76 & 0.491\\
A1656 &12&57&18.3& $+$28 & 12 & 22 & 4000$\div$10000$\ $ & 1.25 & $\ $25 & 100 & $<$0.001\\ 
A2063 &15&20&39.1& $+$08 & 47 & 18 & 9000$\div$12000$\ $ & 0.84 & $\ $11 & $\ $21 & 0.004\\
A2147 &15&59&58.3& $+$16 & 06 & 15 & 8000$\div$14000$\ $ & 0.86 & $\ $21 & $\ $57 & 0.002\\
A2151 &16&02&22.0& $+$17 & 51 & 48 & 8000$\div$14000$\ $ & 0.82 & $\ $25 & $\ $58 & 0.405\\
Pegasus &23&18&00.0& $+$07 & 55 & 00 & 2000$\div$5500$\ $ & 2.36 & $\ $25 & 145 & 0.011\\
\hline
\end{tabular}
\end{center}
{\footnotesize\addtocounter{footnote}{-1}\footnotemark For the Virgo cluster the maximum radial cutoff is $3\ra$.}
\end{table} 

\begin{figure}[t]
\vspace{-0.5cm}
\psfig{figure=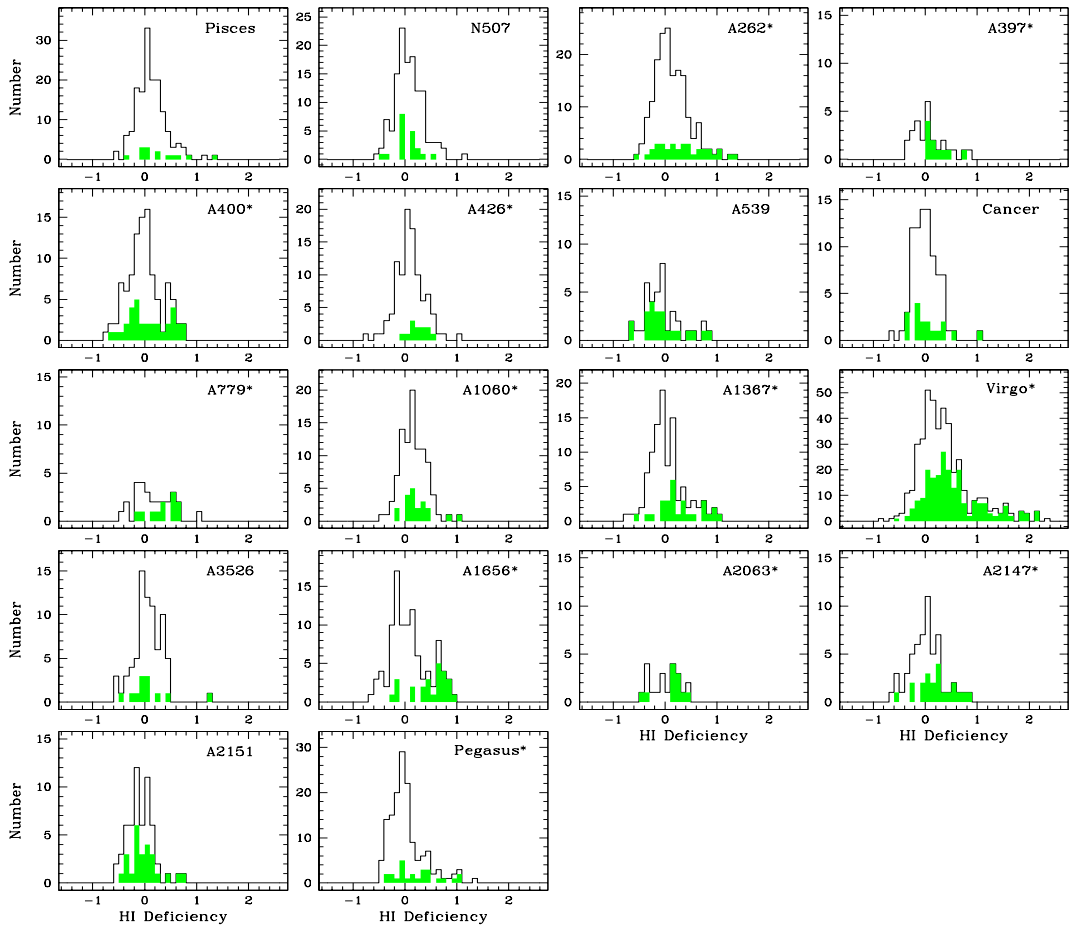,height=14.0cm,width=16.0cm}
\vspace{-0.8cm}
\caption{Histograms showing the distribution of the computed HI
deficiency parameter $\df$ for each cluster region. In each panel the
filled portions of the histogram indicate deficiencies for galaxies
located within $1\ra$ of the cluster center, while the unfilled areas
represent galaxies at larger radii. HI-deficient clusters are
identified by an asterisk after the name.
\label{fig:histo}}
\end{figure}

We have selected from the AGC a total of eighteen cluster fields (see
Table~\ref{tab:tab1}). These are the sky regions centered on the ACO
clusters A262, A397, A400, A426 (Perseus), A539, A779, A1060 (Hydra I),
A1367, A1656 (Coma), A2063, A2147, A2151 (Hercules), and A3526
(Centaurus30), on the clusters of Virgo, Pegasus, Cancer, and Pisces,
and on the group of galaxies around NGC507, hereinafter referred to as
N507. Figure~\ref{fig:histo} shows the histograms of the distribution
of the measured values of $\df$ according to equation~(\ref{def2}) for
these same regions. In the histograms, the filled areas illustrate the
distribution for the galaxies within $1\ra$---for which we adopt the
cluster distance---, while the unfilled areas correspond to those
objects at larger radii. Apart from some expected contamination by
outliers, it is evident from this latter figure that, while the outer
distributions tend to be bell-shaped, exhibit a dispersion comparable
to the value of 0.24 measured for isolated galaxies,\cite{SGH96} and
peak around zero $\df$, the central galaxies of the majority of the
clusters show evidence for strong HI deficiency. The most notable case
is that of Virgo, for which some of the inner galaxies have HI masses
up to two orders of magnitude below their expectation values. Some
samples include a few galaxies undetected in HI but for which a
reliable upper limit of their HI content has been calculated. In the
calculations of $\df$ made through this paper, non-detections always
contribute with their nominal lower limit of deficiency.

We choose to define a cluster as HI deficient when a two-sample
Kolmogorov-Smirnov (KS) test gives a probability of less than 10\% that
its observed inner and outer distributions of $\df$ are drawn from the
same parent population. The results of the KS test confirm essentially
the visual impression: the central spiral populations of twelve cluster
fields, Pegasus, Virgo, A262, A397, A400, A426, A779, A1060, A1367,
A1656, A2063, and A2147 (identified in Fig.~\ref{fig:histo} by an
asterisk after the name) have statistically significant reduced HI
contents. For the other six galaxy concentrations, Cancer, N507,
Pisces, A539, A2151, and A3526, the presence of objects with large HI
deficiencies in the central regions is an exception of the norm.  The
results of the KS test are listed in the last column of
Table~\ref{tab:tab1}.

\begin{figure}[t]
\vspace{-0.5cm}
\psfig{figure=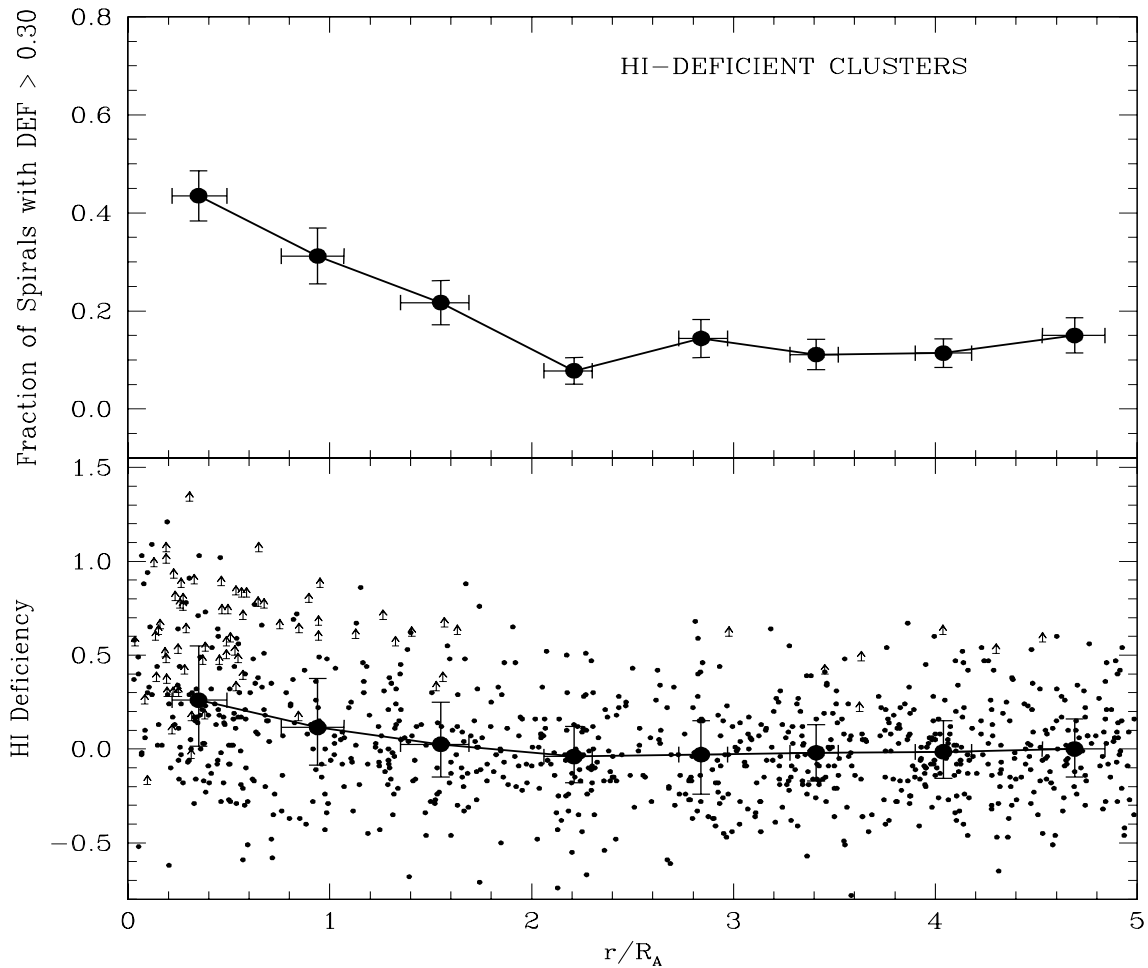,height=14.0cm,width=16.0cm}
\caption{\emph{Top:} HI-deficient fraction in bins of projected radius
from the cluster center for the superposition of all the HI-deficient
clusters but Virgo. \emph{Bottom:} same in upper panel for the measured
HI deficiency. Small dots show the radial variation of HI deficiency
for individual galaxies while the arrows identify non-detections
plotted at their estimated lower limits.
\label{fig:hiradialdef}}
\end{figure}

\begin{figure}[t]
\vspace{-0.5cm}
\psfig{figure=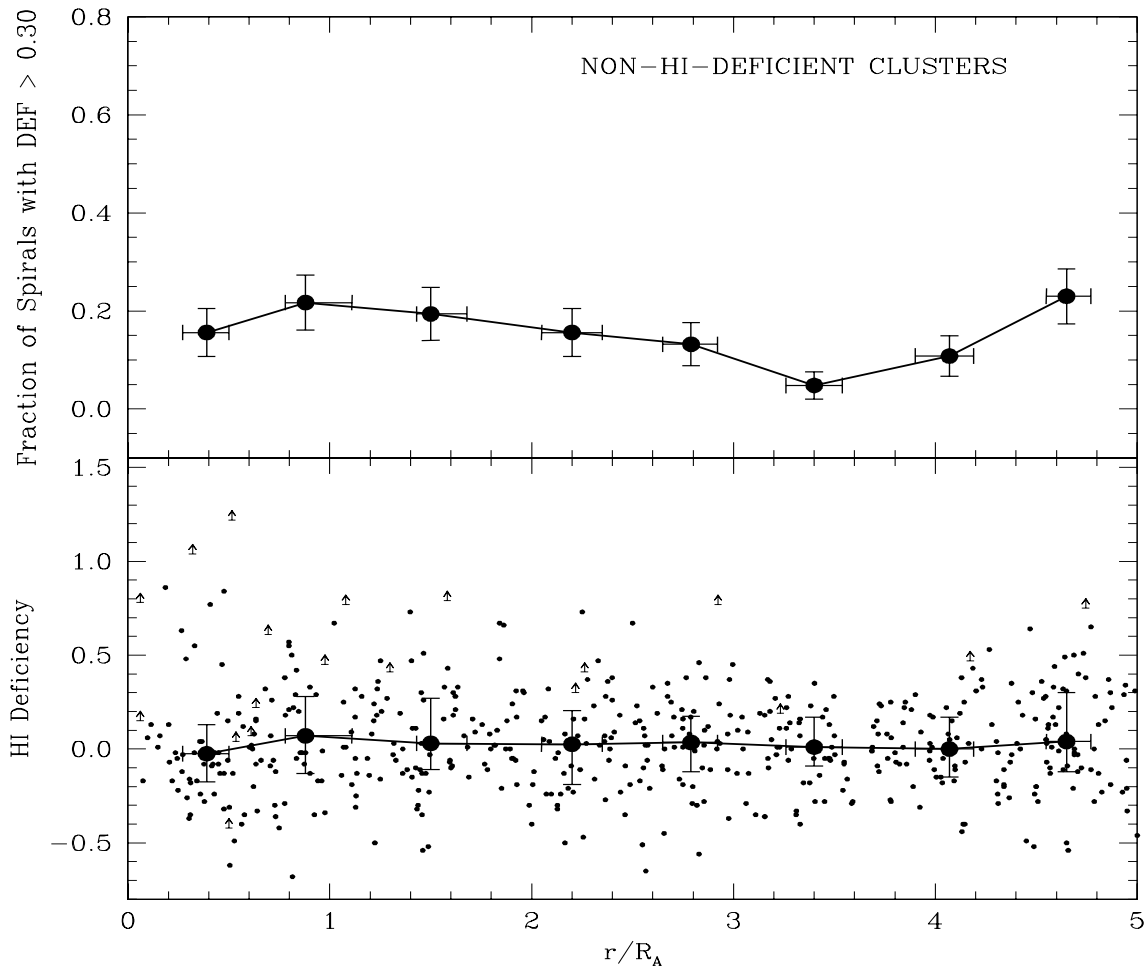,height=14.0cm,width=16.0cm}
\caption{Same as in Figure~\ref{fig:hiradialdef} but for the
superposition of all the non-HI-deficient clusters.
\label{fig:hiradialnodef}}
\end{figure}

\section{The Spatial Pattern of HI Deficiency}\label{pattern}

\subsection{The Radial Variation of HI Deficiency}\label{radial}

A well-known property of the HI deficiency pattern in clusters is its
radial nature. In order to obtain a precise characterization of the
radial behavior of HI deficiency we have combined into a single dataset
the HI measures for spirals in all the clusters which show HI
deficiency other than Virgo, with their clustercentric distances
normalized to $1\ra$. This composite sample of eleven HI-deficient
clusters allows us to trace deficiency out to projected distances from
the cluster center of $5\ra$ in much greater detail than the
(relatively) small samples of the individual clusters, while at the
same time reduces possible distortions caused by substructure and
asphericity. Virgo is excluded from this composite cluster because of
its smaller radial extent and the fact that its much larger dataset
would dominate the composite cluster. To characterize gas removal we
adopt a measure based on the relative populations of deficient and
normal spirals, which has a reduced sensitivity to the presence of
censored data. In Figure~\ref{fig:hiradialdef}, we show the variation
in the fraction of spirals with $\df>0.30$ per bin of projected radial
distance, in Abell radius units, for our composite HI-deficient
cluster. The radial dependence of HI deficiency is clearly evident for
$r < 2\ra$: the percentage of gas-poor spirals increases monotonically
up to the center. Beyond this projected distance, however, the fraction
of gas-deficient disks remains constant around a value of
$\sim$10--15\%, a value consistent with the fraction of field spirals
with $\df>0.30$ expected from a Gaussian distribution of values of this
parameter with an average dispersion of 0.24.~\cite{SGH96} We have also
included in Figure~\ref{fig:hiradialdef} a second panel showing the
variation of HI deficiency with projected radius. It provides visual
verification of the fact that, at large clustercentric distances where
gas-deficient galaxies are scarce and the contribution of
non-detections negligible, the distributions of HI content at different
radii are in excellent agreement both in terms of location and scale
with that of field galaxies. This latter result supports further the
statistical reliability of our measures of $\df$ through
eq.~(\ref{def2}). The same two previous plots are repeated in
Figure~\ref{fig:hiradialnodef} for the superposition of the clusters
which are not deficient in HI.

It is clear from Figure~\ref{fig:hiradialdef} that the influence of the
cluster environment on the neutral gas content of galaxies can extend
beyond one Abell radius (plots of the radial variation of HI deficiency
for individual clusters, not shown here, indicate that the extent of
the zone of significant HI deficiency fluctuates significantly from
cluster to cluster). In this one respect, it is interesting to note
that Balogh {\it et al.},\cite{Bal98} using [OII] equivalent width
data, have found that the mean star formation rate in cluster
galaxies---another property sensitive to environmental effects---shows
signs of depression with respect to the field values at distances
around twice the $R_{200}$ ``virial'' radius. These two results seem to
pose a problem for stripping mechanisms which require high
environmental densities to be effective. The reduction of both the gas
content and the star formation rate of these outer cluster galaxies can
be achieved, however, if they are on strongly eccentric orbits that
have carried them through the cluster center at least once (see
Section~\ref{orbits}). From the theoretical point of view, this
possibility is supported by simulations of hierarchical structure
formation~\cite{Ghi98,RdS98} and by recent direct models of the origin
of clustercentric gradients in the star formation rates within cold
dark matter cosmogonies.~\cite{BNM00}

\begin{figure}[t]
\vspace{-0.8cm}
\psfig{figure=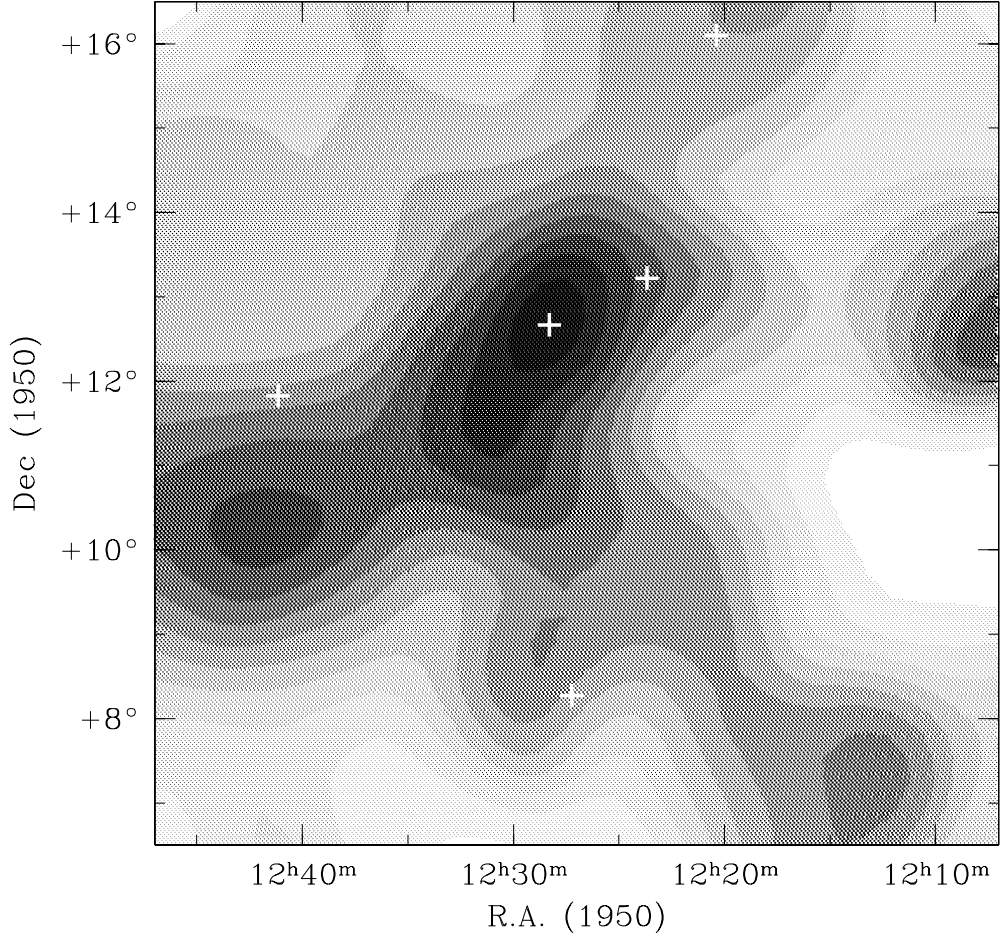,height=16.0cm,width=16.0cm}
\vspace{-1.5cm}
\caption{Greyscale image of the sky distribution of the HI deficiency
in the central region of the Virgo cluster. The lightest regions in the
maps correspond to $\df<0.1$ and the darkest to $\df\ge 1.2$. The
contour spacing is linear. The peak value of HI deficiency coincides
with the maximum of X-ray emission in the whole area. The positions of
five dominant galaxies are marked by crosses (top to bottom: M100, M86,
M87, M60, and M49).
\label{fig:hivirgo}}
\end{figure}

\subsection{The Virgo Cluster}\label{vir3w}

The Virgo cluster is the nearest large-scale galaxy concentration which
offers the possibility of exploring the manifestations of environmental
effects on galaxies with greatest detail. Figure~\ref{fig:hivirgo}
shows the 2D adaptive map of the gas deficiency in the central region
of this cluster (essentially the classical Virgo I Cluster
area). Several structures emerge clearly from the HI deficiency
distribution. The zone with the maximum gas deficiency coincides with
both the peak of X-ray emission and the main density enhancement, known
as Cluster A. This is a double system comprising the subclusters
centered on the giant ellipticals M87 and M86, which seem to be in the
process of merging.~\cite{SBB99} The radial nature of the HI deficiency
pattern can also be observed in the greyscale map of the Virgo core. In
spite of the irregular distribution of the galaxies, the shade
intensity of this map grows toward the position of M87, where the
density of the environment is highest. Five distinct gas-deficient
patches appear to be radially connected with the central one. Two of
them are located along the N-S direction: to the South, the HI
deficiency extends towards the clump dominated by M49 (Cluster B); to
the North, there is a mild increase of gas deficiency around the spiral
M100. Along the EW axis, the distribution of HI deficiency is dominated
by a region of strong gas depletion to the East. This EW asymmetry in
the HI content has also been observed at X-ray wavelengths by
B\"ohringer {\it et al.},\cite{Boh94} who found that the faint Virgo
X-ray emission can be traced out to a distance of $\sim 5\degr$, except
in the western side where the emission falls off more steeply. On the
other hand, the position of the eastern local maximum of deficiency is
located about one and a half degrees South of the peak of the density
enhancement known as Cluster C around the pair of galaxies M59 and M60.
Lumps of high HI-deficient galaxies are also found near the periphery
of the surveyed region, where no X-ray gas is detected, in the areas of
the background galaxy concentrations known as the M cloud in the NW,
and the W' group and (northernmost part of the) W cloud in the SW. We
speculate that these galaxy aggregates, which appear to be connected
with the cluster center by gas-deficient zones, may have already
experienced a first high-velocity passage through the Virgo core that
could have affected the gas content of their galaxies and left behind a
trail of gas-deficient objects, but that would have been insufficient
to tear apart the densest portions of the lumps. It is also remarkable
that the two zones having the lowest gas deficiency in our
HI-deficiency maps---a small region to the East and South of M49 and a
larger one mainly to the South of the M cloud---show a good positional
correspondence with two infalling clouds composed almost entirely
($\sim 80\%$) of spirals.~\cite{Gav99}

\begin{figure}[t]
\vspace{-0.5cm}
\psfig{figure=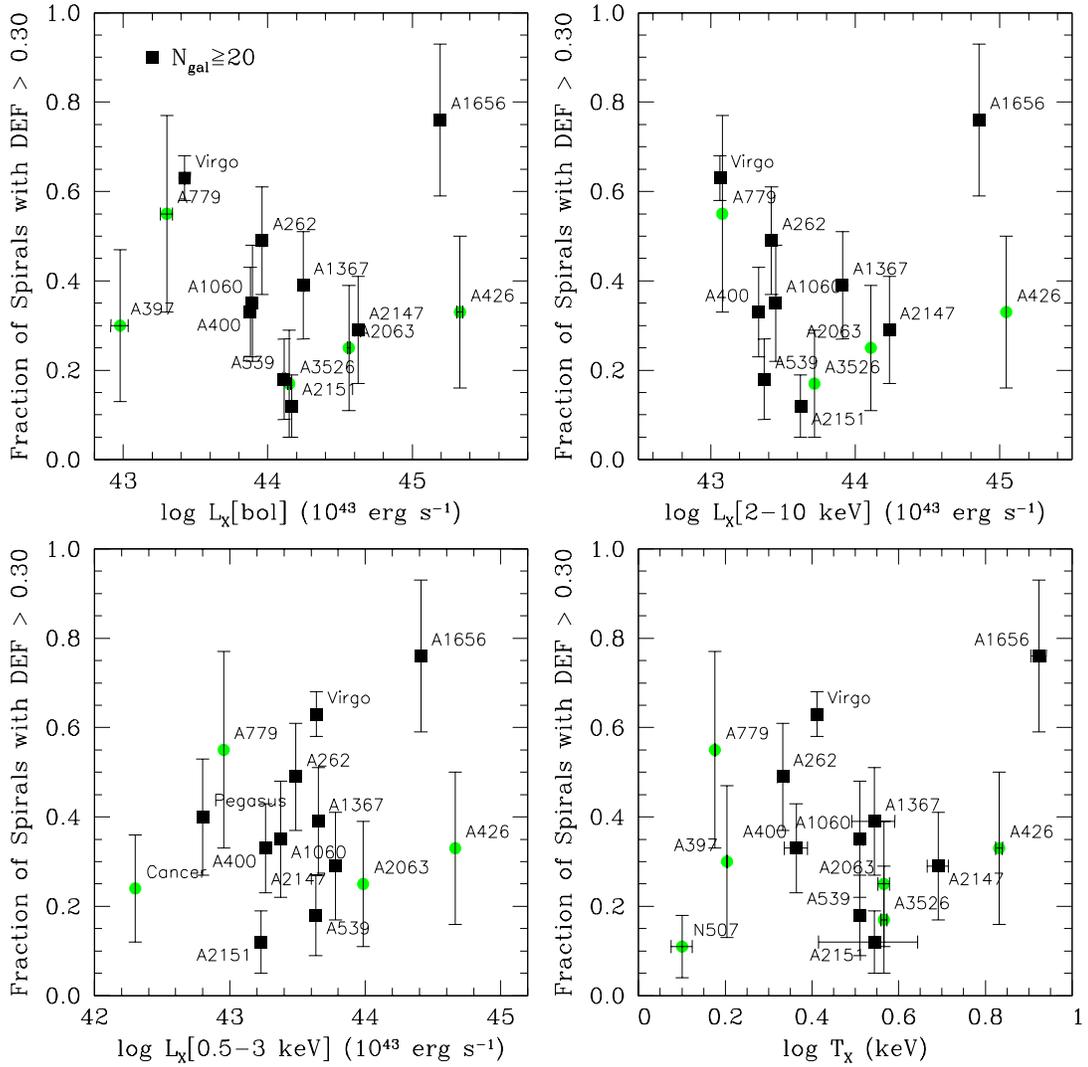,height=16.0cm,width=16.0cm}
\vspace{-0.8cm}
\caption{From left to right and top to
bottom, spiral fraction within $1\ra$ with a deficiency parameter $\df$
larger than 0.30 vs.\ cluster bolometric, 2--10~keV, and 0.5--3.0~keV
X-ray luminosities, and cluster X-ray temperature. Square symbols
identify clusters with a minimum of 20 objects in the central
region. 
\label{fig:fsvsallx}}
\end{figure}

\begin{figure}[t]
\vspace{-0.5cm}
\psfig{figure=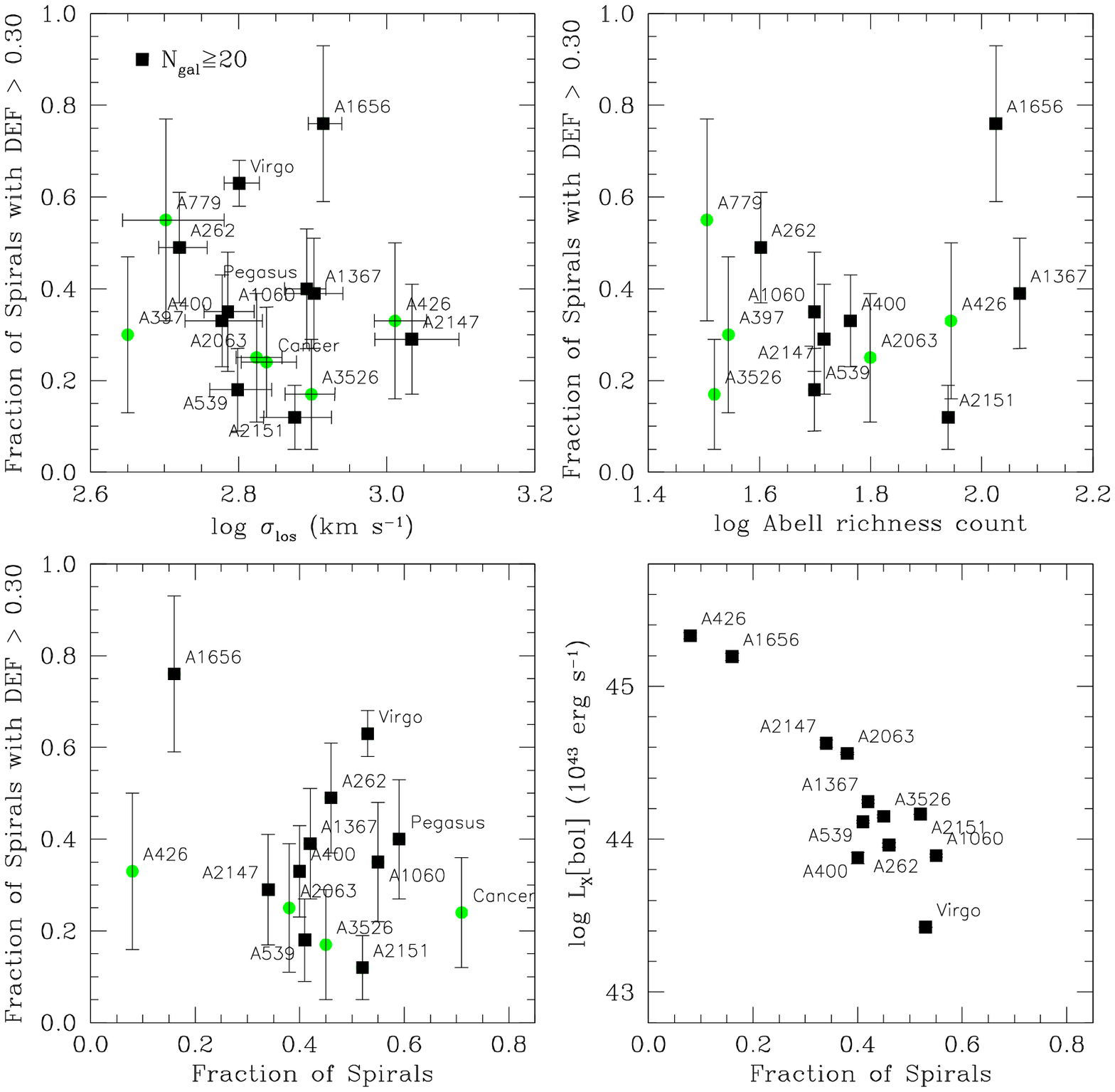,height=16.0cm,width=16.0cm}
\vspace{-0.8cm}
\caption{From left to right and top to
bottom, spiral fraction within $1\ra$ with a deficiency parameter $\df$
larger than 0.30 vs.\ cluster velocity dispersion, Abell richness
count, and total fraction of spirals. The bottom right panel shows the
bolometric X-ray luminosity plotted against the total fraction of
spirals. 
\label{fig:fsvsopt}}
\end{figure}

\section{HI Deficiency and Cluster Properties}\label{clusprop}

The fact that the characteristics of our clusters vary widely suggests
that it is worth investigating correlations between the overall degree
of gas depletion and the global cluster properties that reflect the
strength of the environmental perturbations on the gaseous
disks. Figure~\ref{fig:fsvsallx} shows four X-ray parameters plotted
against the fraction of spirals with $\df>0.30$ found within $1\ra$ of
the cluster center, $\fdef$, while in Figure~\ref{fig:fsvsopt} this
fraction is compared with three optical properties. A fourth panel in
this last figure compares the bolometric X-ray luminosity with the
total spiral fraction. All plots involving the parameter $\fdef$
resemble essentially scatter diagrams with no significant correlations.

We recall at this point that suggestive though inconclusive indications
of a trend toward greater HI-deficient fraction among clusters with
high X-ray luminosity (in the 0.5--3.0~keV band) were found in the
original investigation of Giovanelli and Haynes.~\cite{GH85} This
relationship, however, is not corroborated with the present larger
dataset. One reasonable explanation for the lack of any discernible
correlation is the possible transmutation of some of the swept spirals
into lenticulars, thereby weakening the relationships we are
investigating by reducing the fraction of HI deficient galaxies to a
greater degree for the strong X-ray clusters than for the weak
ones. This possibility is suggested by the very strong anticorrelation
shown by the total spiral fraction and the X-ray luminosity---the linear
correlation coefficient $r$ is less or equal than $-0.90$ for all three
wavebands---in the bottom right panel of Figure~\ref{fig:fsvsopt},
implying that the fraction of lenticulars is correlated with X-ray
luminosity.

One caveat that should be mentioned here is that possible selection
effects and the incompleteness of some of our cluster galaxy samples
might also explain the lack of good correlations between the fraction
of HI-deficient spirals in clusters and the global properties of those
systems. Note, for instance, that X-ray luminous clusters are more
susceptible to incompleteness effects since they have a lower fraction
of spirals. In an attempt to reduce the scatter of the plots, we have
excluded from the analysis those samples containing fewer than than 20
objects, which are the most likely affected by problems related to the
small sample size. Systems in this restricted dataset certainly show
signs of a possible relationship between $\fdef$ and the cluster X-ray
luminosity in the 0.5--3.0~keV range ($r=0.55$), but there is no evidence for
this trend in the other two X-ray windows. We argue that the results of
the present exercise are not fully conclusive and require further
investigation by means of still larger and more complete 21-cm-line
investigation of galaxies in cluster fields.

\begin{figure}[t]
\vspace{-0.5cm}
\psfig{figure=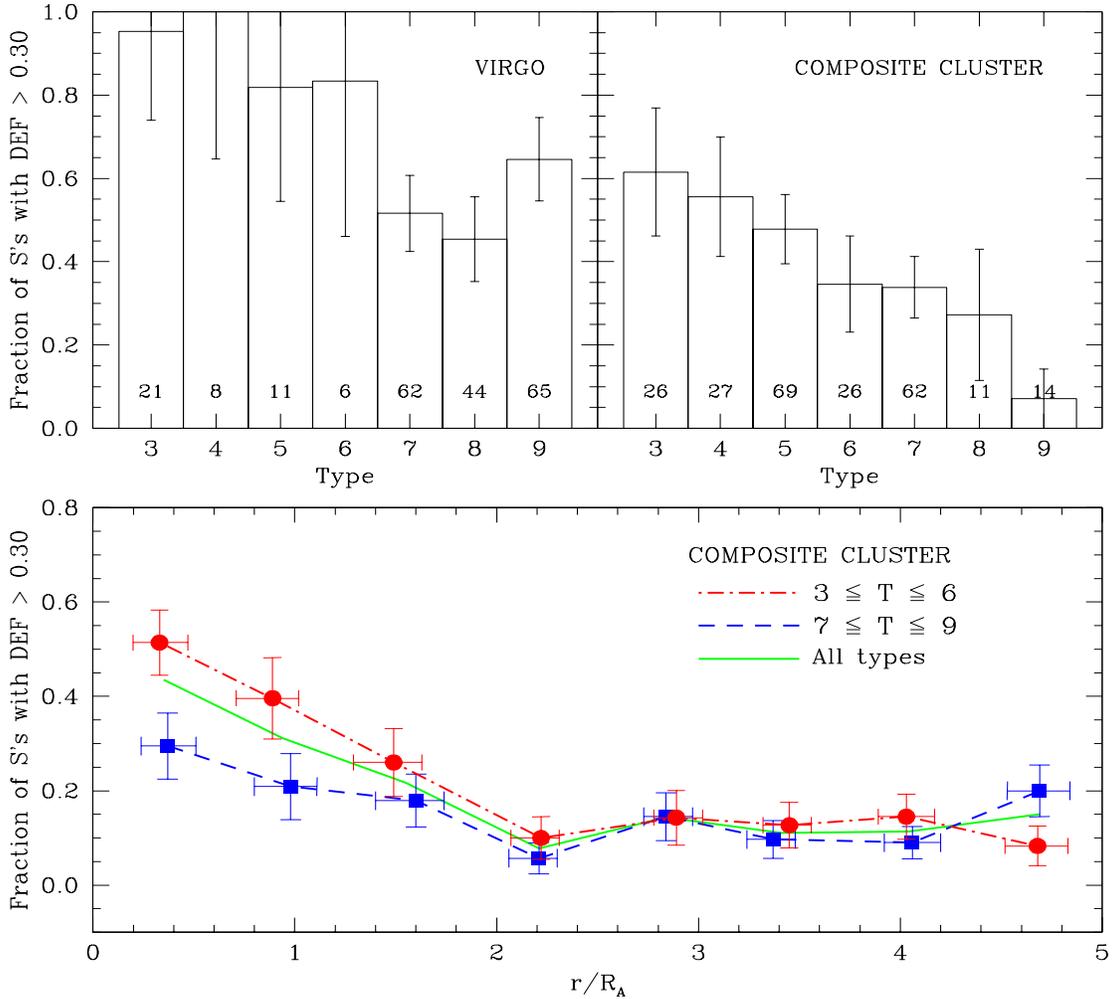,height=14.5cm,width=16.0cm}
\caption{\emph{Top:} fraction of galaxies within $1\ra$ with deficiency
parameter $\df>0.30$ as a function of the morphological type for the
Virgo sample (left) and a composite cluster sample formed by combining
the rest of the HI-deficient systems (right). Hubble types have been
replaced by a numerical code which runs from $T=3$ for Sa's to $T=9$
for Sd--Sm and irregular galaxies. The numbers within each bin indicate
the total number of galaxies in that bin. \emph{Bottom:} same as in the
upper panel of Figure~\ref{fig:hiradialdef}, for the early (circles)
and late (squares) spirals separately. The solid curve reproduces the
trend of the entire spiral population.
\label{fig:type}}
\end{figure}

\section{The Variation of HI Deficiency with Morphological Type}\label{type}

Earlier studies of HI deficient spirals in clusters~\cite{Dre86,CBF86}
have found indications of an increase of the gaseous deficiency towards
early galaxy types. In order to investigate the existence of a gas
deficiency-morphology relationship, we present in the top panels of
Figure~\ref{fig:type} the bar charts of the percentage of galaxies
inside $1\ra$ at a given morphology with deficiency parameter
$\df>0.30$ in the Virgo and the composite samples. Comparison of the
two bar charts shows, in the first place, that the Virgo cluster
exhibits a notably larger fraction of gas-deficient galaxies for any
given morphological class, a result which is simply due to the fact
that we are sampling farther down the HI mass function in Virgo than in
the more distant clusters. Differences in the normalization aside, the
plots confirm that for a spiral, the likelihood of being HI deficient
depends on its morphology. Both the Virgo and the composite cluster
sample share a common pattern: a roughly gradual descent of the
fraction of HI-deficient galaxies as the Hubble type goes from Sa to
Sc, by a total amount of $\sim 40\%$, which levels off for the latest
types. The only discrepancy arises in the very latest morphology bin,
which shows a noticeable recovery of the deficiency fraction for Virgo
and a sharp drop for the composite cluster (we assign more credibility
to the Virgo data since very gas-poor dwarf galaxies are
underrepresented in the more distant clusters). We want to stress also
that the distributions of values of the parameter $\df$ separately by
morphological type demonstrate that the earliest spirals, as well as the
dwarfs, are not only more likely to be deficient in HI than the
intermediate disks, but they also have a higher gas deficiency.

On the other hand, the bottom panel of Figure~\ref{fig:type}
demonstrates that the observed correlation between HI deficiency and
morphology is not simply a consequence of the well-known morphological
segregation of cluster galaxies. In this plot, we reproduce the radial
run of the HI-deficient fraction for the composite HI-deficient cluster
sample but separating the early ($T$:\,$3\div 6$) and late
($T$:\,$7\div 9$) spiral type subsets. From this graph one sees: 1)
that inside the region of influence of the cluster environment,
early-type galaxies have systematically higher gas deficiencies,
\emph{at any projected radius}, than the late types; and 2) that the
difference between the HI-deficient fractions of the early- and
late-type populations increases gradually towards the cluster
center. Identical results, although with more abrupt radial variations
due to the spatial lumpiness of the cluster, are found for the Virgo
sample. These results lead to the conclusion that the observed
correlation of HI deficiency and morphology is not a secondary effect
of the spatial segregation of the galaxies, but reflects the interplay
between the intrinsic characteristics of these objects and the physical
mechanism behind HI depletion.

\section{HI Deficiency and Galaxy Orbits}\label{orbits}

The hypothesis that HI-deficient galaxies lose their interstellar HI at
small distances from cluster cores but can still be found at large
radial distances suggests that the HI-deficient objects follow highly
eccentric orbits. Information on the eccentricity of galaxy orbits can
be extracted from the radial run of the line-of-sight (los) velocity
dispersion.

\begin{figure}[t]
\vspace{-0.5cm}
\psfig{figure=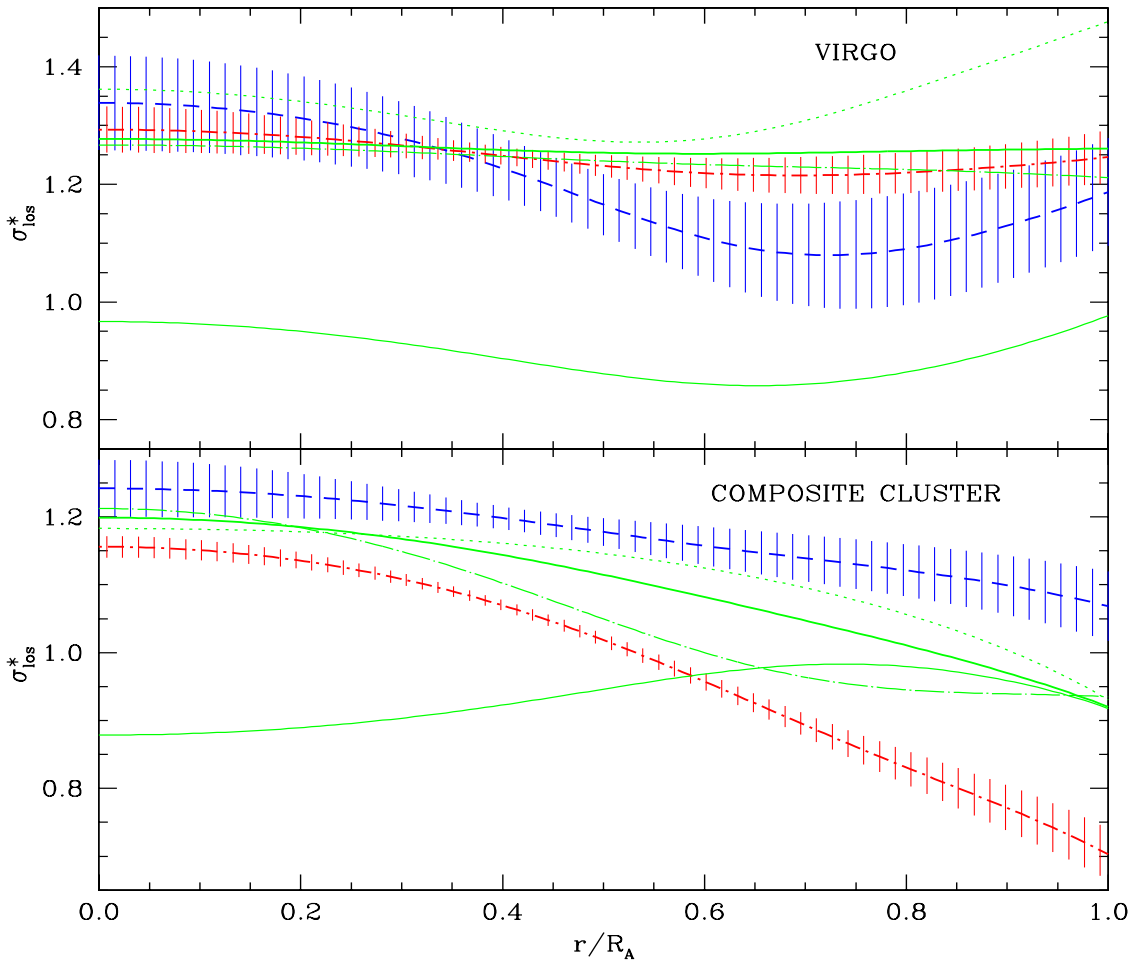,height=14.0cm,width=16.0cm}
\caption{Radial run of the \emph{normalized} los velocity dispersion up
to $1\ra$ for the Virgo (top) and the composite HI-deficient cluster
(bottom). The observed peculiar velocities of the individual galaxies
have been scaled to the average los velocity dispersion of their parent
cluster. Line coding is as follows: thick dot-dash for spirals with
$\df\ge 0.48$, thick dash for spirals with $\df\le 0$, thick solid for
all spirals, dots for early spirals, dot-long dash for late spirals,
and solid for ellipticals and lenticulars. Only error bars for the
curves of the spirals with extremal HI contents are displayed for
clarity.
\label{fig:vlos}}
\end{figure}

In Figure~\ref{fig:vlos}, we show, separately for the Virgo and
composite clusters, the mean radial profiles of the normalized los
velocity dispersion, $\sigma^\ast_\mathrm{los}$, for six different
galaxian subpopulations. For the composite HI-deficient cluster, we see
that the velocity dispersion for the spirals with the strongest gas
deficiencies ($\df\ge 0.48$) drops significantly in a manner consistent
with radial orbits.~\cite{Dre86} The curve for the gas-rich objects
($\df\le 0$) decreases too with increasing radius, although the decline
is sensibly weaker than for the gas-deficient galaxies. These results
suggest that one possible explanation for the relationship found in the
previous section between disk morphology and gas content could be that
early spirals have an orbital distribution more radially anisotropic
than late types. To test this possibility, we have inferred the
velocity dispersion profiles of the spirals subdivided into early and
late disks. Again, we find indications of radial orbits for these two
broad morphological groupings, although the trajectories of the
galaxies in the first group does not seem to be more eccentric than
those in the second: if anything there is a hint for the opposite
effect. Not surprisingly, the kinematic behavior of the entire spiral
population is intermediate among those shown by all the previous
subdivisions. On the other hand, the earliest Hubble types, i.e.\
lenticulars and ellipticals, show a lower and almost constant velocity
dispersion profile compatible with an isotropic distribution of
velocities. This finding is yet another manifestation of the well-known
fact that S and E+S0 galaxies do not share the same kinematics:
late-type galaxies are likely recent arrivals to the virialized cluster
cores, which consist essentially of ellipticals and
lenticulars.~\cite{Sod89} In addition to supporting this basic picture,
our data also indicate that a segregation has developed among the
orbits of the infalling spirals according to their gaseous contents
since the objects with the more eccentric trajectories,
\emph{regardless of morphology}, reach deeper into the cluster cores
and are thus more efficiently stripped of their neutral hydrogen.

The same analysis for the Virgo cluster galaxies is reproduced in the
top panel of Figure~\ref{fig:vlos}. We see that, to a first
approximation, the velocity dispersion profiles corresponding to all
the galaxy subgroups are essentially flat (notice, for instance, the
curve exhibited by the entire spiral population), although with a
noticeably positive excess of the velocity dispersion of the spirals
relative to the E+S0 population. Given that Virgo is a dynamically
young galaxy system, we interpret these results as indicative of the
fact that the trajectories of the spiral galaxies within the central
region of this cluster are strongly perturbed by large and rapid
fluctuations of the mean gravitational field caused by the ongoing
merger of major subclumps. Because of this large-scale phase mixing,
environmental influences on the disks have not yet been capable of
inducing a neat orbital segregation between gas-poor and gas-rich
objects. The presence of high-velocity gas-poor objects at relatively
large clustercentric distances would result from the spirals that
populate the outskirts of the infalling clouds and that, still bound to
these systems, have been scattered with high velocities to large
apocenter orbits during the merger process.

\section{Implications of the Results on the Mechanism of HI 
Depletion}\label{discussion}

The natural consequence of gas losses as radical as our results
indicate would be---provided gas replenishment does not occur at
exceptionally high rates---a reduction in the star-formation activity
of the galaxy, followed by the fading of the disk and the consequent
increase of the bulge to disk ratio (B/D). This prediction is in good
agreement with the decline of disk luminosity and the invariance of
bulge brightness with increasing local density observed in the spirals
of rich clusters.~\cite{SSS89} Also quite consistent with this idea is
the finding by Koopmann and Kenney~\cite{KK98} that objects in Virgo
classified as Sa have similar B/D's than the Sc's and only differ in
their overall star formation rates which are strongly reduced in the
outer disk. The morphological transformation of the swept galaxies into
S0-like objects could be completed through the suppression of the
spiral features by continued disk heating by tidal encounters. Of
course, the possible morphological evolution of cluster spirals towards
earlier types has serious difficulties in explaining the presence of
S0's in the field. While some stripped galaxies may have fairly radial
orbits that carry them at large distances from the cluster centers, one
must bear in mind that not all the lenticular galaxies, outside and
inside clusters, arise necessarily from HI-deficient spirals.

The present investigation provides clear evidences of the strong
influence that the cluster environment has on the gaseous disks of
spirals. The marked radial pattern of HI deficiency indicates that
galaxies lose their gas near the cluster centers. This result is
consistent with the finding that spirals with substantial HI deficiency
follow orbits with large radial components. It appears then, that the
stripping of gas requires high IGM densities and relative
velocities. The detection of galaxies with extreme HI deficiencies, but
still retaining their spiral morphology, suggests also that the
stripping of the atomic hydrogen is a relatively recent event in the
life of these objects. Furthermore, the details of the relationship
between HI deficiency and morphology are consistent with the idea that
some galaxian characteristics directly related to the Hubble type, such
as for instance the presence of central depressions in the HI disks,
enhance the efficiency of gas removal. According to these results,
ISM-IGM interactions, basically ram pressure supplemented by the
accompanying effects of viscosity and turbulence, are favored over
other environmental mechanisms as the main cause of gas depletion in
clusters.

\end{document}